\documentclass[aip,jmp,amsmath,amssymb,
preprint,
]{revtex4-1}

\usepackage{graphicx}
\usepackage{dcolumn}
\usepackage{bm}
\usepackage[utf8]{inputenc}
\usepackage[T1]{fontenc}
\usepackage{mathptmx}
\usepackage{etoolbox}
\usepackage[hidelinks]{hyperref}

\hypersetup{
    colorlinks=true,
    linkcolor=blue
    }

\newcommand{\Det}{\mathrm{Det}}
\newcommand{\Tr}{\mathrm{Tr}}
\newcommand{\sgn}{\mathrm{sgn}}

\makeatletter
\def\@email#1#2{%
 \endgroup
 \patchcmd{\titleblock@produce}
  {\frontmatter@RRAPformat}
  {\frontmatter@RRAPformat{\produce@RRAP{*#1\href{mailto:#2}{#2}}}\frontmatter@RRAPformat}
  {}{}
}%
\makeatother
\begin{document}

%\preprint{}

\title[Exact solution of the Wegner flow equation]{The exact solution of the Wegner flow equation with the Mielke generator for $3\times 3$ Hermitian matrices}
\author{Tomasz Masłowski}
 \altaffiliation{The Faculty of Mathematics and Applied Physics, Rzeszów University of Technology, \\ al.~Powstańców Warszawy 6, 35-959 Rzeszów, Poland}
 \email{tmaslowski@prz.edu.pl}

\date{\today}

\begin{abstract}
 The exact solution of the Wegner flow equation with the Mielke generator for $3\times 3$ Hermitian matrices is presented.
  The general solutions for $N\times N$ tridiagonal Hermitian matrices and partially for $4\times 4$ real symmetric matrices are also given.
\end{abstract}

\maketitle

\section{\label{sec:level1}Introduction}

The Wegner flow equation \cite{Wegner1994} is a continuous unitary transformation leading to a near-diagonal structure of a Hamiltonian.
A similar equation was earlier analysed by Chu and Driessel \cite{Chu1990} and Brockett \cite{BROCKETT1991} but not in a physical context.
Another example of unitary transformation that brings Hamiltonian closer to diagonalization
is Głazek and Wilson similarity renormalization group \cite{Glazek1994} which introduces more
freedom in designing the connection of the initial (bare) Hamiltonian and the effective one.

The Wegner flow equation has been applied to various physical problems in many-body systems \cite{Kehrein2006}, Anderson localization \cite{Quito2016}, dissipative systems \cite{Rosso2020,Schmiedinghoff2022}, QED \cite{Gubankova1998,Cetin2023}, QCD \cite{Glazek2012,Gomez2015} and many others, see e.g. \cite{Wegner2006} as a review.

The unitary transformation of a Hermitian operator (Hamiltonian), $H$, may be expressed by means of the transformation generator, $G$,
\begin{equation}
 \frac{d H(s)}{d s} = \left[G(s),H(s)\right] \;,\label{general}
\end{equation}
with the initial condition
\begin{equation}
H(s=0)=H_0 \;.
\end{equation}
The solution to this equation gives the whole family of Hamiltonians, $H(s)$, parametrized by $s$ and can be considered as a continuous unitary flow.
In the case of original Wegner flow equation the generator is chosen as the commutator of an operator with its diagonal part, i.e.:
\begin{equation}
G_W(s)=\left[D(s),H(s)\right] \;, \label{wg}
\end{equation}
where $D(s)$ denotes the diagonal part of $H(s)$. 

Mielke in \cite{Mielke1998} presented an alternative choice of generator.
It is given by a skew-Hermitian matrix built form $H(s)$. We will denote the set of Hermitian matrices of the size $N\times N$ as $\mathrm{Herm}(N, \mathbb{C})$ and the set of real symmetric matrices of size $N\times N$ as $\mathrm{Sym}(N, \mathbb{R})$. For $H\in \mathrm{Herm}(3, \mathbb{C})$
\begin{equation}
H(s) = \begin{bmatrix}
a(s) & b(s) & g(s) \\
b^*(s) & d(s) & c(s) \\
g^*(s) & c^*(s) & f(s) 
  \end{bmatrix}\;,
\end{equation}
the Mielke generator is
\begin{equation}
G_M(s)= \begin{bmatrix}
0 & b(s) & g(s) \\
-b^*(s) & 0 & c(s) \\
-g^*(s) & -c^*(s) & 0 
  \end{bmatrix}\;.
\end{equation}

Regardless of the generator, the Wegner flow equation leads to a set of ordinary nonlinear differential equations. The Wegner generator leads to the third degree equation in matrix elements of $H(s)$, while the Mielke generator to the second degree. Thus, we choose the Mielke generator for the sake of simplicity. Another choice leading the Wegner equation to be quadratic in matrix elements could be a generator of Brockett type, i.e. given by $[F,H(s)]$, where $F$ is a constant diagonal matrix.

As the Wegner flow equation as well as the Głazek-Wilson renormalization group were designed for perturbative calculation, there are not too many attempts in search for exact solutions. We can include in this group the proof of existence of limit cycles \cite{Glazek2002}, the calculation of spectrum in a specially designed model Hamiltonian \cite{Glazek2021} and results in the spin-boson model in the Ohmic bath \cite{Mielke2000}.

The paper is organized in the following way. In Sec. \ref{sol3} we derive the solution for
$\mathrm{Sym}(3, \mathbb{R})$, Sec.~\ref{herm} generalizes the result
to any $H\in\mathrm{Herm}(N, \mathbb{C})$, Sec. \ref{NxN} presents
the general solution for $N\times N$ tridiagonal matrices, Sec. \ref{c4x4} presents the partial solution in the case of $H\in\mathrm{Sym}(4, \mathbb{R})$, Sec~\ref{eigen}
shows the connection between the obtained solution and the eigenvalue problem of $H_0$ and we summarize in Sec. \ref{discuss}. In addition there are 5 appendices
containing details and examples.

\section{\label{sol3}$3\times 3$ real symmetric matrix}

The solution to (\ref{general}) for $H\in \mathrm{Sym}(2, \mathbb{C})$ presents no difficulties
for majority of generators of interest, including $G_W$, $G_M$ or the Brockett type, though it leads to non-trivial physical results, e.g. \cite{Rosso2020,Schmiedinghoff2022}.
For the Mielke generator the solution is given in App. \ref{c2x2}.
The first mathematically demanding case is for $H\in \mathrm{Sym}(3, \mathbb{R})$, but still the solution
presented here requires only basic notions of linear algebra and ordinary differential equations. 
First, we solve (\ref{general}) with the Mielke generator for $H\in\mathrm{Sym}(3, \mathbb{R})$ and in the next section generalize to $H\in\mathrm{Herm}(3, \mathbb{C})$.
For $H\in\mathrm{Sym}(3, \mathbb{R})$  (\ref{general}) leads to % (the argument $s$ is omitted for better reading):
\begin{eqnarray}
  a'&=&2\left(b^2+g^2\right) \;,  \label{seta} \\
  d'&=&-2\left(b^2-c^2\right)\;,  \label{setd} \\
  f'&=&-2\left(c^2+g^2\right)\;,  \label{setf} \\
  b'&=&-b\,(a-d)+2cg \;, \label{setb} \\
  c'&=&-c\,(a-d)-2bg \;, \label{setc} \\
  g'&=&-g\,(a-d) \;. \label{setg} 
\end{eqnarray}  
To construct the solution we introduce new variables:
\begin{eqnarray}
  z(s)=a(s)-d(s)\;, \label{defz} \\
  t(s)=d(s)-f(s)\;,  \label{deft} \\
  \beta_1(s)=b^2(s)\;,  \label{defb1} \\
  \beta_2(s)=c^2(s)\;,  \label{defb2} \\
  \gamma(s)=g^2(s)\;. \label{defg} 
\end{eqnarray}
As the flow equation describes the unitary evolution $\Tr(H)$ is conserved, then the functions $a(s)$, $d(s)$ and $f(s)$ are linearly
dependent and one of them may be eliminated. Thus, using new variables we deal only with 5 equations:
\begin{eqnarray}
  \phantom{w1111}z' &=& 2\left(2\beta_1-\beta_2+\gamma\right)\;, \label{z} \\
  \phantom{w1111}t' &=& 2\left(-\beta_1+2\beta_2+\gamma\right)\;, \label{t} \\
  \left(\log \beta_1\right)' &=& -2z+4\sqrt{\frac{\beta_2\gamma}{\beta_1}}\;, \label{b1} \\
  \left(\log \beta_2\right)' &=& -2t-4\sqrt{\frac{\beta_1\gamma}{\beta_2}} \;, \label{b2} \\
  \left(\log \gamma\right)' &=& -2(z+t)\;. \label{b3} 
\end{eqnarray}

\subsection{\label{simc}Simplified case}

We start the construction of solution for $g_0=0$ which gives $g(s)=0$ (this is not true for $G_W$).
This simplify (\ref{z}) -- (\ref{b2}) to
\begin{eqnarray}
  \phantom{w1111}z' &=& 2\left(2\beta_1-\beta_2\right)\;, \label{zs} \\
  \phantom{w1111}t' &=& 2\left(-\beta_1+2\beta_2\right)\;, \label{ts} \\
  \left(\log \beta_1\right)' &=& -2z\;, \label{b1sps} \\
  \left(\log \beta_2\right)' &=& -2t \label{b2sps} 
\end{eqnarray}
Eliminating $z$ and $t$ we get a closed set of equations for $\beta_1$ and $\beta_2$:
\begin{eqnarray}
  \left(\log \beta_1\right)'' &=& 4\left(-2\beta_1+\beta_2\right)\;, \label{b1pps} \\
  \left(\log \beta_2\right)'' &=& 4\left(\beta_1-2\beta_2\right) \;. \label{b2pps}
\end{eqnarray}
As for any positive functions $\zeta_i$ and real numbers $\delta_i$ there holds the identity
\begin{equation}
  \left(\log (\zeta_1^{\delta_1}\zeta_2^{\delta_2}\cdot\ldots )\right)'' = \delta_1\frac{\widetilde{\zeta}_1}{\zeta_1^2} + \delta_2\frac{\widetilde{\zeta}_2}{\zeta_2^2} + \ldots \;,\label{logdd}
\end{equation}
where we denote
\begin{equation}
  \widetilde{\zeta} \equiv \zeta\zeta''- \left(\zeta'\right)^2\;,
\end{equation}
then all factors in $\beta_1$ and $\beta_2$ must be presented on the RHS of (\ref{b1pps}) and (\ref{b2pps}) as denominators in $\beta_1$ or $\beta_2$.
This observation allows us to look for the solution in the form
\begin{eqnarray}
  \beta_1 &=& \frac{\eta_2}{\eta_1^2} \;, \label{b1s} \\
  \beta_2 &=& \frac{\eta_1}{\eta_2^2} \;. \label{b2s}
\end{eqnarray}
With this choice  (\ref{b1pps}) and (\ref{b2pps}) are fulfilled if $\eta_1$ and $\eta_2$ satisfy
\begin{eqnarray}
  4\eta_1 &=& \widetilde{\eta}_2 \;, \label{ce1s} \\
  4\eta_2 &=& \widetilde{\eta}_1 \;. \label{ce2s}
\end{eqnarray}
By analogy with the $2\times 2$ case, see App. \ref{c2x2}, we look for $\eta_i$ in the form
\begin{eqnarray}
\eta_1 &=& p^{(1)}_1 \,e^{u_1 s} + p^{(1)}_2 \,e^{u_2 s} + p^{(1)}_3\,e^{u_3 s} \;,  \label{e1t}\\
\eta_2 &=& p^{(2)}_1 \,e^{v_1 s} + p^{(2)}_2 \,e^{v_2 s} + p^{(2)}_3\,e^{v_3 s} \;.  \label{e2t}
\end{eqnarray}
Comparison of factors of exponential functions in  (\ref{ce1s}) and (\ref{ce2s}) leads to
\begin{eqnarray}
  0 &=& u_1+u_2+u_3  \;, \label{w0} \\
  0 &=& v_1+v_2+v_3 \;, \label{v0} \\
  v_i &=& -u_i \;, \label{vw} \\
  p^{(2)}_1 &=& \frac{1}{4}\,p^{(1)}_2\,p^{(1)}_3\left(u_2-u_3\right)^2\;,\quad \mathrm{cycl.} \label{psimp} \;,\\
  p^{(1)}_1 &=& \frac{1}{4}\,p^{(2)}_2\,p^{(2)}_3\left(u_2-u_3\right)^2\;,\quad \mathrm{cycl.} \label{psimp2} 
\end{eqnarray}
The above equations are overdetermined, then they lead to the following constraint
(we consider here only different $u_i$, otherwise see App. \ref{appA6})
\begin{equation}
p^{(1)}_1\, p^{(1)}_2\, p^{(1)}_3 = p^{(2)}_1\, p^{(2)}_2\, p^{(2)}_3 = \frac{64}{\left(u_1-u_2\right)^2 \left(u_1-u_3\right)^2 \left(u_2-u_3\right)^2} \;. \label{p64} 
\end{equation}

Furthermore, since $\beta_1$ and $\beta_2$ are positive, then 
$\eta_1$ and $\eta_2$ are positive as well, thus $p^{(1)}_i$ being the coefficient of $e^{u_i s}$ with the highest $u_i$ must be positive, the same is for $p^{(2)}_i$. Taking into account  (\ref{psimp}) and  (\ref{psimp2}) yields that all $p^{(1)}_i$ and $p^{(2)}_i$ must be positive. This ends the construction of the solution to  (\ref{b1pps}) and (\ref{b2pps}).
The solutions for $z$ and $t$ come straightforwardly from  (\ref{b1sps}) and (\ref{b2sps})
\begin{eqnarray}
  z &=& \frac{\eta_1'}{\eta_1} - \frac{\eta_2'}{2\eta_2} \;, \label{zf}\\
  t &=& -\frac{\eta_1'}{2\eta_1} + \frac{\eta_2'}{\eta_2} \;. \label{tf} 
\end{eqnarray}
Summarizing, the solution to  (\ref{zs}) -- (\ref{b2sps}) and then to (\ref{seta}) -- (\ref{setc}) with $g_0=0$ can be expressed by two functions $\eta_1$ and $\eta_2$ given by (\ref{e1t}) -- (\ref{e2t}), with parameters satisfying  (\ref{w0}) -- (\ref{p64}) and all $p^{(i)}_j$ being positive.
The connection of parameters $p^{(i)}_j$ and $u_i$ with $H_0$ entries is presented in Sec. \ref{icon}.

\subsection{\label{gsimc}General symmetric case}

Once the simplified case is solved the solution of $g_0\ne 0$ case can be found in a simple way.
 (\ref{z}) and (\ref{t}) give
\begin{eqnarray}
  \beta_1 &=& \frac{1}{6}(2z'+t') - \gamma \;, \\
  \beta_2 &=& \frac{1}{6}(z'+2t') - \gamma \;.
\end{eqnarray}
If we assume the form of $z$ and $t$ from the previous case, i.e. given by  (\ref{zf}) and (\ref{tf}),
then  (\ref{b3}) gives 
\begin{equation}
  \gamma = \frac{1}{\eta_1\eta_2} \;. \label{b3f} 
\end{equation}
This allows to express $\beta_1$ and $\beta_2$ in terms of $\eta_1$ and $\eta_2$
\begin{eqnarray}
  \beta_1 &=&  \frac{\widetilde{\eta}_1}{4\eta_1^2} - \frac{1}{\eta_1\eta_2} \;, \label{b1f} \\
  \beta_2 &=&  \frac{\widetilde{\eta}_2}{4\eta_2^2} - \frac{1}{\eta_1\eta_2}\;. \label{b2f} 
\end{eqnarray}
Since $b(s)$ and $c(s)$ are square roots of $\beta_1$ and $\beta_2$, and $b(s)$ and $c(s)$ may change a sign, then there should exist functions $\rho_1$ and $\rho_2$ such that
\begin{eqnarray}
  \beta_1 &=&  \frac{\rho_1^2}{4\eta_1^2\eta_2} \;, \label{b1rho} \\
  \beta_2 &=&  \frac{\rho_2^2}{4\eta_1\eta_2^2} \label{b2rho} 
\end{eqnarray}
and
\begin{eqnarray}
  \rho_1^2 &=& \widetilde{\eta}_1\eta_2 -4 \eta_1 \;, \label{ro1-e} \\
  \rho_2^2 &=& \widetilde{\eta}_2\eta_1 -4 \eta_2 \;. \label{ro2-e} 
\end{eqnarray}
Let us assume $\eta_i$ in the form obtained for $g_0=0$, i.e.:
\begin{eqnarray}
\eta_1(s) &=& p^{(1)}_1 \,e^{u_1 s} + p^{(1)}_2 \,e^{u_2 s} + p^{(1)}_3 \,e^{-(u_1+u_2) s} \;,  \label{e1tf}\\
\eta_2(s) &=& p^{(2)}_1 \,e^{-u_1 s} + p^{(2)}_2 \,e^{-u_2 s} + p^{(2)}_3 \,e^{(u_1+u_2) s} \label{e2tf}
\end{eqnarray}
and $\rho_1$ and $\rho_2$ in the form
\begin{eqnarray}
\rho_1(s) &=& q^{(1)}_1 \,e^{-u_1 s} + q^{(1)}_2 \,e^{-u_2 s} + q^{(1)}_3 \,e^{(u_1+u_2) s} \;,  \label{rho1t}\\
\rho_2(s) &=& q^{(2)}_1 \,e^{u_1 s} + q^{(2)}_2 \,e^{u_2 s} + q^{(2)}_3 \,e^{-(u_1+u_2) s} \;.  \label{rho2t}
\end{eqnarray}
Then (\ref{ro1-e}) and (\ref{ro2-e}) can be solved for $q^{(i)}_j$ by comparison of coefficients of
exponential functions, see App. \ref{appR}.

Finally, we can express the solution to the Wegner equation in terms of $\eta_1$, $\eta_2$ and dependent on them $\rho_1$ and $\rho_2$:
\begin{eqnarray}
  a(s) &=& \frac{\Tr(H_0)}{3} + \frac{\eta_1'(s)}{2\eta_1(s)} \;, \label{aa}\\
  d(s) &=& \frac{\Tr(H_0)}{3} - \frac{\eta_1'(s)}{2\eta_1(s)} + \frac{\eta_2'(s)}{2\eta_2(s)} \;,\\
  f(s) &=& \frac{\Tr(H_0)}{3} - \frac{\eta_2'(s)}{2\eta_2(s)} \;,\label{ff} \\
  b(s) &=& \frac{\rho_1(s)}{2\,\eta_1(s) \sqrt{\eta_2(s)}} \;,\label{bb}\\
  c(s) &=& \frac{\rho_2(s)}{2\sqrt{\eta_1(s)}\, \eta_2(s)} \;,\label{cc} \\
  g(s) &=& \pm\frac{1}{\sqrt{\eta_1(s) \eta_2(s)}} \;, \label{gg}
\end{eqnarray}
where the sign in (\ref{gg}) is equal the $g_0$ sign.

Although, we used only  (\ref{z}), (\ref{t}), (\ref{b3}) and the trace conservation,
the resulting functions $\beta_{1,2}$ satisfy also  (\ref{b1}) and (\ref{b2}). This is due to that the unitary (similarity) transformation introduces more dependencies between parameters of functions constituting the system (\ref{z}) -- (\ref{b3}) than we needed to analyze.
The proof that (\ref{aa}) -- (\ref{gg}) are really the solution which covers all possibilities is given in the next section where we express parameters of $\eta_1$ and $\eta_2$ by entries of $H_0$.

\subsection{\label{icon}Initial conditions}

The initial conditions were already partially taken into account by the use of the condition that $\Tr(H)$ is conserved.
Let us note that the parameters $p^{(i)}_j$ in $\eta_{1,2}$ are ambiguous as $z$ and $t$ contain terms  $\eta_i'/\eta_i$. Thus, we redefine $p^{(1)}_i$ and $p^{(2)}_i$ as $p^{(1)}_i/p^{(1)}_3$ and $p^{(2)}_i/p^{(2)}_3$ and call $p^{(1)}_3 p^{(2)}_3$ as $A^{-2}$. Doing this $\eta_i$ take the form
\begin{eqnarray}
\eta_1 &=& p^{(1)}_1 \,e^{u_1 s} + p^{(1)}_2 \,e^{u_2 s} + e^{-(u_1+u_2) s} \;,  \label{e1s}\\
\eta_2 &=& p^{(2)}_1 \,e^{-u_1 s} + p^{(2)}_2 \,e^{-u_2 s} + e^{(u_1+u_2) s} \;,  \label{e2s}
\end{eqnarray}
Changing the normalization of $\eta_1(s)$ and $\eta_2(s)$ , i.e. values of $\eta_1(0)$ and $\eta_2(0)$, does not affect functions $a(s)$, $d(s)$ and $f(s)$, while
$\rho_1(s)$ and $\rho_2(s)$ which define $b(s)$ and $c(s)$, need to be corrected, the same with $g(s)$.
This is because definitions  (\ref{ro1-e}) and (\ref{ro2-e}) are not homogeneous in $\eta_i$.
If $\eta_1(s)$ and $\eta_2(s)$ are both normalized to $1/|g_0|$ then $\rho_1(s)$ and $\rho_2(s)$ are given by  (\ref{ro1-e}) and (\ref{ro2-e}), and $g(s)$ by
 (\ref{gg}).
If $\eta_1(s)$ and $\eta_2(s)$ are given by  (\ref{e1s}) and (\ref{e2s}) then $g(s)$ is given by
\begin{equation}
g(s) = \frac{\sgn(g_0)A}{\sqrt{\eta_1(s) \eta_2(s)}} \;, \label{ggA}
\end{equation}
and instead of  (\ref{ro1-e}) and (\ref{ro2-e}) we have
\begin{eqnarray}
  \rho_1^2 &=& \widetilde{\eta}_1\eta_2 -4A \eta_1 \;, \label{ro1-eA} \\
  \rho_2^2 &=& \widetilde{\eta}_2\eta_1 -4A \eta_2 \;. \label{ro2-eA} 
\end{eqnarray}
In the case of other normalizations in the place of $A$ a proper factor have to appear.
 
Thus, we have 7 unknowns: $u_1$, $u_2$, $p^{(1)}_1$, $p^{(1)}_2$, $p^{(2)}_1$, $p^{(2)}_2$ and $A$.
To find $u_1$ and $u_2$ we consider the limit when $s\rightarrow\infty$. When $s\rightarrow\infty$, then $H(s)$ tends to the diagonal form with its eigenvalues on the diagonal.
Additional feature of the flow equations is that the eigenvalues are sorted \cite{BROCKETT1991}.
With our choice of $G_M$ the largest eigenvalue is $a(\infty)=a_\infty$ and the smallest $f(\infty)=f_\infty$. As for any $x_k, y_k \in \mathbb{R}$
\begin{equation}
\lim_{s\rightarrow\infty} \frac{\left(\sum_k x_k e^{y_k s}\right)'}{\sum_k x_k e^{y_k s}} = \max_k y_k \;, \label{presort}
\end{equation}
then, see (\ref{aa}) -- (\ref{ff}),
\begin{eqnarray}
  a_\infty &=& \frac{\Tr(H_0)}{3} + \frac{u_1}{2} \;, \label{ainf}\\
  f_\infty &=& \frac{\Tr(H_0)}{3} - \frac{u_2}{2} \;, \label{finf}
\end{eqnarray}
where we have chosen $u_1=\max(u_1,u_2)$ and $u_2=\min(u_1,u_2)$.\\
$a_\infty$, $f_\infty$ and $d_\infty=d(\infty)$ are the solution of the eigenvalue equation for $H_0$, i.e.:
\begin{equation}
x^3-I_1\, x^2 -I_2\, x - I_3 =0 \;, \label{cubic}
\end{equation}
where ($I_k$ are principal invariants of $H_0$)
\begin{eqnarray}
  I_1 &=& \Tr(H_0) \;,\\
  I_2 &=& \frac{1}{2}\left(\Tr^2(H_0)-\Tr(H_0^2)\right) \;,\\
  I_3 &=& \Det(H_0) \;.
\end{eqnarray}
It is more convenient to consider the depressed cubic equation 
\begin{equation}
x^3+P x +Q =0 \label{depq3}
\end{equation}
where
\begin{eqnarray}
  P &=& - \frac{1}{3} \left( I_1^2+3\,I_2\right)\; , \label{PP} \\
  Q &=& - \frac{1}{27} \left( 2\,I_1^3+9\,I_1I_2+27\,I_3\right)\;. \label{QQ}
\end{eqnarray}
Taking into account  (\ref{ainf}) and (\ref{finf}) means that $u_1/2$, $u_2/2$ and $-(u_1+u_2)/2$ are the solutions to the depressed cubic equation.
As all roots are real then the set of roots to  (\ref{depq3}) can be written as:
\begin{widetext}
\begin{equation}
\mathcal{R}=\left\{2\sqrt{-\frac{P}{3}} \cos\left(\frac{1}{3} \arccos\left(\frac{3 Q}{2 P} \sqrt{-\frac{P}{3}}\right) - \frac{2 \pi k}{3}\right)\bigg| k\in\{0,1,2\}\right\} \label{R3}
\end{equation}
\end{widetext}
then
\begin{eqnarray}
  u_1 &=& 2\max \mathcal{R} \;, \label{u1} \\
  u_2 &=& 2\min \mathcal{R} \;, \label{u2}
\end{eqnarray}
or alternatively
\begin{equation}
u_i=2\left(x_i-\frac{\Tr(H_0)}{3}\right) \;, \label{r2u}
\end{equation}
where
$\{x_1,x_2,x_3\}\equiv \mathcal{R}$.
Thus, all exponents in $\eta_1$, $\eta_2$, $\rho_1$ and $\rho_2$ are known.\\
To find the rest of 5 coefficients, $p^{(1)}_1$, $p^{(1)}_2$, $p^{(2)}_1$, $p^{(2)}_2$ and $A$ we use the initial conditions written as
\begin{eqnarray}
  a_0 &=& a(0) \;, \label{a0}\\
  f_0 &=& f(0) \;, \label{f0}\\
  b_0^2 &=& \beta_1(0) \;, \\
  c_0^2 &=& \beta_2(0) \;, \\
  g_0^2 &=& \gamma(0) \;.
\end{eqnarray}
First, from the first four equations we find $p^{(i)}_j$ in terms of $A$ and then the fifth equation gives $A^2$.
Finally, we get (for $u_1+2u_2= 0$ or $2u_1+u_2= 0$ see App. \ref{appA6}):
\begin{eqnarray}
  A &=&  \frac{9\left| (u_1 +2u_2) (2u_1 + u_2)\right|}{\sqrt{D_1D_2}}\, |g_0| \;, \label{A}\\
  p^{(1)}_1 &=& \frac{(u_1 +2 u_2)N_{11}}{(u_1 - u_2)D_1} \;, \label{p11}\\
  p^{(1)}_2 &=& \frac{(2u_1 + u_2)N_{12}}{(u_1 - u_2)D_1} \;,\\
  p^{(2)}_1 &=& \frac{(u_1 +2 u_2)N_{21}}{(u_1 - u_2)D_2} \;,\\
  p^{(2)}_2 &=& \frac{(2u_1 + u_2)N_{22}}{(u_1 - u_2)D_2} \;,\label{p22}
\end{eqnarray}
where
\begin{widetext} 
\begin{eqnarray}
  D_1 &=& 36 a_0^2 + 36 b_0^2 + 36 g_0^2 + 6 u_1 I_1 + 4 I_1^2 + 9 u_1 u_2 + 6 I_1 u_2 - 6 a_0 (3 u_1 + 4 I_1 + 3 u_2) \;, \label{D1}\\
  D_2 &=& 36 c_0^2 + 36 f_0^2 + 36 g_0^2 + 6 u_1 I_1 + 4 I_1^2 + 9 u_1 u_2 + 6 I_1 u_2 - 6 f_0 (3 u_1 + 4 I_1 + 3 u_2)\;, \label{D2}\\
  N_{11} &=&  36 a_0^2 + 36 b_0^2 + 36 g_0^2 - 24 a_0 I_1 + 4 I_1^2 - 9 u_1 u_2 + 18 a_0 u_1 - 6 I_1 u_1 - 9 u_2^2\;,\\
  N_{12} &=& -36 a_0^2 - 36 b_0^2 - 36 g_0^2 + 24 a_0 I_1 - 4 I_1^2 + 9 u_1 u_2 - 18 a_0 u_2 + 6 I_1 u_2 + 9 u_1^2\;,\\
  N_{21} &=&  36 f_0^2 + 36 c_0^2 + 36 g_0^2 - 24 f_0 I_1 + 4 I_1^2 - 9 u_1 u_2 + 18 f_0 u_1 - 6 I_1 u_1 - 9 u_2^2\;,\\
  N_{22} &=& -36 f_0^2 - 36 c_0^2 - 36 g_0^2 + 24 f_0 I_1 - 4 I_1^2 + 9 u_1 u_2 - 18 f_0 u_2 + 6 I_1 u_2 + 9 u_1^2\;. \label{N22}
\end{eqnarray}
\end{widetext}
This ends the proof that (\ref{aa}) -- (\ref{gg}) describe all solutions to  (\ref{general}).

\section{\label{herm}$3\times 3$ Hermitian matrix}

For $H\in\mathrm{Herm}(3, \mathbb{C})$ we use the notation
\begin{equation}
H(s) = \begin{bmatrix}
a(s) & b(s)\,e^{i\varphi_b(s)} & g(s)\,e^{i\varphi_g(s)} \\
b(s)\,e^{-i\varphi_b(s)} & d(s) & c(s)\,e^{i\varphi_c(s)} \\
g(s)\,e^{-i\varphi_g(s)} & c(s)\,e^{-i\varphi_c(s)} & f(s) 
  \end{bmatrix}\;.
\end{equation}
Now, functions $b(s)$, $c(s)$ and $g(s)$ represent moduli of appropriate matrix elements and as such
cannot be negative. We also use variables defined in (\ref{defz}) -- (\ref{defg}).
The Wegner equation now leads to
\begin{eqnarray}
  \left(\log \beta_1\right)' &=& -2z+4\sqrt{\frac{\beta_2\gamma}{\beta_1}}\cos\left(\varphi_g-\varphi_b-\varphi_c \right)\;, \label{hb1} \\
  \left(\log \beta_2\right)' &=& -2t-4\sqrt{\frac{\beta_1\gamma}{\beta_2}}\cos\left(\varphi_g-\varphi_b-\varphi_c \right) \;, \label{hb2} 
\end{eqnarray}
while equations for $z'(s)$, $t'(s)$ and $\gamma'(s)$ retain their forms, i.e.  (\ref{z}), (\ref{t}) and (\ref{b3}) are still valid.
In addition we have 3 new equations for $\varphi_b(s)$, $\varphi_c(s)$ and $\varphi_g(s)$
\begin{eqnarray}
  \varphi_b' &=& 2\sqrt{\frac{\beta_2\gamma}{\beta_1}}\sin\left(\varphi_g-\varphi_b-\varphi_c \right)\;, \label{hphib1} \\
  \varphi_c' &=& -2\sqrt{\frac{\beta_1\gamma}{\beta_2}}\sin\left(\varphi_g-\varphi_b-\varphi_c \right) \;, \label{hphib2} \\
  \varphi_g' &=& 0\;. \label{hphig} 
\end{eqnarray}

The first step to find the solution to the system of  (\ref{z}), (\ref{t}), (\ref{b3}), (\ref{hb1}) -- (\ref{hphig}) is
to recognise that as  (\ref{z}), (\ref{t}) and (\ref{b3}) remain unchanged we can repeat the same reasoning as for $\mathrm{Sym}(3, \mathbb{R})$.
This brings us to the same formulas for $z(s)$, $t(s)$, $\beta_1(s)$, $\beta_2(s)$ and $\gamma(s)$. Thus,  (\ref{zf}), (\ref{tf}) and (\ref{b3f}) -- (\ref{b2f}) as well as $A$ and parameters of $\eta_i(s)$ given by (\ref{A}) -- (\ref{N22}) are also valid now. At this point similarities between symmetric and Hermitian
cases end. For a Hermitian matrix functions $b(s)$ and $c(s)$ represent moduli, therefore they cannot change a sign, then we do not demand the existence of functions $\rho_1$ and $\rho_2$ from  (\ref{b1rho}) -- (\ref{ro2-e}). Thus, $b(s)$ and $c(s)$ are given by
\begin{eqnarray}
  b(s) &=& \pm \frac{\sqrt{\widetilde{\eta}_1(s)\eta_2(s) -4 \eta_1(s)}}{2\eta_1(s)\sqrt{\eta_2(s)}} \;, \label{bherm} \\
  c(s) &=& \pm \frac{\sqrt{\widetilde{\eta}_2(s)\eta_1(s) -4 \eta_2(s)}}{2\sqrt{\eta_1(s)}\eta_2(s)} \;, \label{cherm} 
\end{eqnarray}
where signs of $b(s)$ and $c(s)$ are sings of $b_0$ and $c_0$.\\
Furthermore, 
even if the symmetric and Hermitian matrices differ only by phases $\varphi_{b0}$, $\varphi_{c0}$ and $\varphi_{g0}$, then functions $\eta_i(s)$ in the Hermitian case are not the same as for the symmetric case, though in both cases we have the same set of equations
and the same initial conditions for $a(s)$, $d(s)$, $f(s)$, $b(s)$ $c(s)$ and $g(s)$. This is because phases $\varphi_i$ cause that these two matrices have different eigenvalues. This makes that exponential functions in $\eta_i(s)$ have different exponents as well as all other parameters, as they depend on exponents.

Once we have $z(s)$, $t(s)$, $\beta_1(s)$, $\beta_2(s)$ and $\gamma(s)$ we can use  (\ref{hb1}) -- (\ref{hphig})
to find $\varphi_b(s)$, $\varphi_c(s)$ and $\varphi_g(s)$. Equation (\ref{hphig}) gives
\begin{equation}
  \varphi_g(s) = \varphi_{g0} \;. \label{hphib2sol}
\end{equation}
To separate equations for $\varphi_b(s)$ and $\varphi_c(s)$ we add  (\ref{hb1}) and (\ref{hb2}) and subtract (\ref{b3}). This gives
\begin{equation}
  \left(\log \frac{\beta_1\beta_2}{\gamma}\right)'= 4\left(\sqrt{\frac{\beta_2\gamma}{\beta_1}}-\sqrt{\frac{\beta_1\gamma}{\beta_2}}\right)
  \cos\left(\varphi_g-\varphi_b-\varphi_c \right)\;,
\end{equation}
while adding  (\ref{hphib1}) and (\ref{hphib2}) gives
\begin{equation}
  \left(\varphi_b+\varphi_c\right)'= 2\left(\sqrt{\frac{\beta_2\gamma}{\beta_1}}-\sqrt{\frac{\beta_1\gamma}{\beta_2}}\right)
  \sin\left(\varphi_g-\varphi_b-\varphi_c \right)\;.
\end{equation}
Thus,
\begin{widetext}
\begin{equation}
  \left(\log \frac{\beta_1\beta_2}{\gamma}\right)'= 2\cot\left(\varphi_g-\varphi_b-\varphi_c \right)\left(\varphi_b+\varphi_c\right)'
  = -2\left[\log \sin \left(\varphi_g-\varphi_b-\varphi_c \right)\right]' \;,
\end{equation}
\end{widetext}
then
\begin{equation}
  \frac{\beta_1\beta_2}{\gamma} \sin^2 \left(\varphi_g-\varphi_b-\varphi_c \right) = C^2 \;, \label{CC}
\end{equation}
where $C$ is a constant. Its value can be easily determined,
\begin{equation}
  C = \frac{b_0 c_0}{g_0} \sin \left(\varphi_{g0}-\varphi_{b0}-\varphi_{c0} \right) \;.
\end{equation}
Finally, by substituting  (\ref{CC}) to  (\ref{hphib1}) and (\ref{hphib2}) we get (see also the note at the beginning of Sec. \ref{icon})
\begin{eqnarray}
  \varphi_b' &=& 2C\frac{\gamma}{\beta_1} = 8C\frac{\eta_1}{\widetilde{\eta}_1\eta_2 -4 \eta_1} \;, \label{hphib1p} \\
  \varphi_c' &=& -2C\frac{\gamma}{\beta_2} = -8C\frac{\eta_2}{\widetilde{\eta}_2\eta_1 -4 \eta_2}   \label{hphib2p} 
\end{eqnarray}
and
\begin{eqnarray}
  \varphi_b(s) &=& \varphi_{b0} + 8C\int_0^s \frac{\eta_1(s')}{\widetilde{\eta}_1(s')\eta_2(s') -4 \eta_1(s')}ds'  \;, \label{hphib1i} \\
  \varphi_c(s) &=& \varphi_{c0} -8C\int_0^s \frac{\eta_2}{\widetilde{\eta}_2(s')\eta_1(s') -4 \eta_2(s')}ds'\;.  \label{hphib2i} 
\end{eqnarray}
Thus, $\varphi_b(s)$ and $\varphi_c(s)$ as well as all matrix elements of $H(s)$ are fully described by $\eta_1(s)$ and $\eta_2(s)$. If $g_0=0$, then
\begin{eqnarray}
  \varphi_b(s) &=& \varphi_{b0} \;, \\
  \varphi_c(s) &=& \varphi_{c0} \;.
\end{eqnarray}

App. \ref{appE1} illustrates both cases, symmetric and Hermitian, with examples.

\section{\label{NxN}$N\times N$ tridiagonal matrix}

In this section we derive a general solution for the $N\times N$ tridiagonal Hermitian matrix,
\begin{widetext}
\begin{equation}
H(s) = \begin{bmatrix}
a_1(s) & b_1(s)e^{i\varphi_1(s)} & 0 & \cdots & 0\\
b_1(s)e^{-i\varphi_1(s)} & a_2(s) & b_2(s)e^{i\varphi_2(s)} & & \vdots  \\
0 & b_2(s)e^{-i\varphi_1(s)} & \ddots & \ddots & 0 \\
\vdots & &\ddots &  & b_{N-1}(s)e^{i\varphi_{N-1}(s)} \\
0 &\cdots & 0 & b_{N-1}(s)e^{-i\varphi_{N-1}(s)} & a_N(s)\\
  \end{bmatrix}\;.
\end{equation}
\end{widetext}
In analogy to  (\ref{defz}) -- (\ref{defg}) we introduce new variables:
\begin{eqnarray}
  z_k(s) &=& a_k(s) - a_{k+1}(s)\;, \\
  \beta_k(s) &=& b_k^2(s) \;,
\end{eqnarray}
where $k=1,\ldots,N-1$.\\
In these variables  (\ref{general}) reads as
\begin{eqnarray}
  \phantom{WW} z_1' &=& 2\left(2 \beta_1 - \beta_2\right) \;, \label{ztri1} \\
  \phantom{WW} z_2' &=& 2\left(- \beta_1 +2\beta_2 - \beta_3\right)\;, \\
  &\vdots&  \nonumber \\
  \phantom{kk} z_{N-1}' &=& 2\left(- \beta_{N-2} +2\beta_{N-1}\right)\;, \\
  \left(\log \beta_k\right)' &=& -2 z_k \;, \quad k=1,\ldots,N-1 \label{ztriN-1} \;,\\
  \phantom{WW} \varphi_k'&=& 0  \;, \quad k=1,\ldots,N-1 \label{phiN-1} \;.
\end{eqnarray}
The last equation gives
\begin{equation}
  \varphi_k(s) = \varphi_{k0}  \;, \quad k=1,\ldots,N-1 \label{phiN-1sol} \;.
\end{equation}
Thus, for tridiagonal matrices the hermiticity does not introduce more complications than symmetricity.
The rest of equations leads to
\begin{eqnarray}
  \phantom{kl} \left(\log \beta_1\right)'' &=& 4 \left(-2 \beta_1 + \beta_2\right) \;, \label{bt1} \\
  \phantom{kl} \left(\log \beta_2\right)'' &=& 4 \left(\beta_1 -2 \beta_2 + \beta_3\right) \;, \label{bt2}\\
  &\vdots&  \nonumber \\
  \left(\log \beta_{N-1}\right)'' &=& 4 \left(\beta_{N-2} -2 \beta_{N-1}\right) \;. \label{btN-1}
\end{eqnarray}
Analogically to the $3\times3$ case we look for the solution to the above system of equations in the form:
\begin{eqnarray}
  \beta_1 &=& \frac{\eta_2}{\eta_1^2} \;, \label{betatri1} \\
  \beta_2 &=& \frac{\eta_1\eta_3}{\eta_2^2} \;, \label{betatri2} \\
  &\vdots&  \nonumber \\
  \beta_k &=& \frac{\eta_{k-1}\eta_{k+1}}{\eta_k^2} \;, \label{betatrik} \\
  &\vdots&  \nonumber \\
  \beta_{N-1} &=& \frac{\eta_{N-2}}{\eta_{N-1}^2} \;.  \label{betatriN-1}
\end{eqnarray}
what leads to conditions functions $\eta_i$ have to satisfy
\begin{eqnarray}
  \widetilde\eta_1&=&4\eta_2 \;, \label{tild-beta1}\\
  \widetilde\eta_2&=&4\eta_1\eta_3 \;,\label{tild-beta2}\\
  &\vdots&  \nonumber \\
  \widetilde\eta_k&=&4\eta_{k-1}\eta_{k+1} \;,\label{tild-betak}\\
  &\vdots&  \nonumber \\
  \widetilde\eta_{N-1}&=&4\eta_{N-2} \;. \label{tild-betaN-1}
\end{eqnarray}
Unlike it is for $\mathrm{Herm}(3,\mathbb{C})$ the above equations cannot be solved in terms
of functions $\eta_i$ with a fixed number of exponential functions. Fortunately, 
we can solve  (\ref{tild-beta1}) -- (\ref{tild-betaN-1}) recursively.
Assuming 
\begin{equation}
  \eta_1(s) = \sum_{i=1}^N p^{(1)}_i \,e^{u_i s} \;, \label{eta_1_gen}
\end{equation}
we find that
\begin{equation}
  \eta_k(s) = \sum_{\{i\}\in U^k_N} p^{(k)}_{\{i\}}\,e^{\sum_{j\in \{i\}} u_j s} \;, \label{eta_k}
\end{equation}
where in analogy to  (\ref{r2u})
\begin{equation}
u_i=2\left(w_i-\frac{\Tr(H_0)}{N}\right) \label{eigen2exp}
\end{equation}
and $w_i$ are eigenvalues of $H_0$, $U_N=\{1,2,\ldots, N\}$ and $U^k_N$ is a set of all $k$-combinations of $U_N$ elements.
Thus, $\eta_k$ consists of $\binom{N}{k}$ exponential functions what means that for $N\ge 4$ only $\eta_1$ and $\eta_{N-1}$ are built from
$N$ exponential functions, while all other $\eta_k$ have them more than $N$. For a given $N$ the set of all $\eta_k$ we denote by $\mathcal{E}^N_k$.
Let us note that  (\ref{eigen2exp}) gives
\begin{equation}
\sum_{i=1}^N u_i= 0 \label{sumu}\;.
\end{equation}
First we check that the relation (\ref{eigen2exp}) gives a correct relationship
between eigenvalues of $H_0$ and exponents of exponential functions in $\eta_i$.
Similarly to (\ref{aa}) -- (\ref{ff}) we assume the formula for $a_i(s)$ as
\begin{eqnarray}
  a_1(s)&=&\frac{\Tr(H_0)}{N} + \frac{\eta_1'(s)}{2\eta_1(s)}\;, \label{aN41} \\
  a_2(s)&=&\frac{\Tr(H_0)}{N} - \frac{\eta_1'(s)}{2\eta_1(s)} + \frac{\eta_2'(s)}{2\eta_2(s)}\;,\\
  &\vdots& \nonumber \\
  a_{N-1}(s)&=&\frac{\Tr(H_0)}{N} - \frac{\eta_{N-2}'(s)}{2\eta_{N-2}(s)} + \frac{\eta_{N-1}'(s)}{2\eta_{N-1}(s)}\;,\\
  a_N(s)&=&\frac{\Tr(H_0)}{N} - \frac{\eta_{N-1}'(s)}{2\eta_{N-1}(s)}  \label{aN4N}
\end{eqnarray}
Applying (\ref{presort}) to (\ref{aN41}) -- (\ref{aN4N}) we get
\begin{eqnarray}
  a_1(\infty)&=&\frac{\Tr(H_0)}{N} + \frac{u_1}{2} = w_1\;, \label{iu1} \\
  a_2(\infty)&=&\frac{\Tr(H_0)}{N} - \frac{u_1}{2} + \frac{u_1+u_2}{2} = w_2 \;,\\
  &\vdots& \nonumber \\
  a_{N-1}(\infty)&=&\frac{\Tr(H_0)}{N} - \frac{u_1+\ldots+u_{N-2}}{2} + \frac{u_1+\ldots+u_{N-1}}{2}=w_{N-1}\;,\\
  a_N(\infty)&=&\frac{\Tr(H_0)}{N} - \frac{u_1+\ldots+u_{N-1}}{2} = w_N \;, \label{iuN}
\end{eqnarray}
where the second equalities come from (\ref{eigen2exp}) and  (\ref{iuN}) is granted from (\ref{sumu}). We denote the largest $w_i$ as $w_1$, the second largest as $w_2$, etc.
Then, the set $\{w_1,w_2,\cdots,w_N\}$ is sorted descendingly and $\{u_1,u_2,\cdots,u_N\}$ does the same. This shows the validity of assumption (\ref{eigen2exp}). In addition we can see how is realized the sorting mechanism proven in \cite{BROCKETT1991}.

To find relations between coefficients $p^{(k)}_i$, for $k=2,\ldots,N-1$, we express them in terms of $p^{(1)}_i$ and $u_i$.
Comparing coefficients of exponential functions in  (\ref{tild-beta1}) we obtain
\begin{equation}
p^{(2)}_{ij} = \frac{1}{4} p^{(1)}_i p^{(1)}_j \left(u_i-u_j\right)^2 \;. \label{p2sol}
\end{equation}
Next, (\ref{tild-beta2}) gives the relationship between $p^{(3)}_{ijk}$ and $p^{(2)}_{ij}$, but 
$p^{(2)}_{ij}$ can be eliminated using (\ref{p2sol}), then we get
\begin{equation}
p^{(3)}_{ijk} = \frac{1}{4^3} p^{(1)}_i p^{(1)}_j p^{(1)}_k \left(u_i-u_j\right)^2 \left(u_i-u_k\right)^2 \left(u_j-u_k\right)^2 \;. \label{p3sol}
\end{equation}
Repeating this we express all parameters of $\eta_k$ where $k\ge 2$, ending with
\begin{widetext}
\begin{equation}
p^{(N-1)}_{ij\ldots k} = \frac{1}{2^{(N-1)(N-2)}} p^{(1)}_i p^{(1)}_j \cdot \ldots \cdot p^{(1)}_k \left(u_i-u_j\right)^2 \cdot \ldots \cdot \left(u_i-u_k\right)^2 \cdot \ldots \cdot \left(u_j-u_k\right)^2 \;. \label{pNsol}
\end{equation}
\end{widetext}
The same could be done by starting with $\eta_{N-1}$ instead of $\eta_1$ what leads to the constrain on $p^{(1)}_i$ being a generalization of  (\ref{p64}) (we limit here to non-degenerate $H_0$)
\begin{equation}
  \prod_{i=1}^N p^{(1)}_i = \frac{2^{N(N-1)}}{\prod_{1\le i<j\le N} \left(u_i-u_j\right)^2 } \;, \label{FF}
\end{equation}
which comes out when considering the factor of all $p^{(N-1)}_i$ expressed by $p^{(1)}_i$ and vice versa.

Finally, we get the general solution to the Wegner equation expressed in terms of $N$ parameters $p^{(1)}_i$ which all are bound together by the constraint (\ref{FF}), this gives $N-1$ free parameters. The tridiagonal $H_0$ has $2N-1$ entries. $N$ of them can be fixed using $N$ exponents $u_i$, then $N-1$ are left. This agrees with the number of free $p^{(1)}_i$. As there are no general formula for the solution of quintic equation and above in terms of radicals, the connection of $p^{(1)}_i$ with the initial condition represented by $H_0$ would be explicitly possible only in special cases. In App. \ref{appE2} we present the example of $5\times 5$ matrix.

\section{\label{c4x4} $4\times 4$ symmetric matrix}

In this section we present the sketch of the construction of solution for $H\in \mathrm{Sym}(4, \mathbb{R})$. 
We follow the same line of reasoning as in the section \ref{sol3}.
As previously, for a given 
\begin{equation}
H(s) = \begin{bmatrix}
a_1(s) & b_1(s) & d_1(s) & g(s)   \\
b_1(s) & a_2(s) & b_2(s) & d_2(s) \\
d_1(s) & b_2(s) & a_3(s) & b_3(s) \\
g(s) & d_2(s) & b_3(s) & a_4(s)
  \end{bmatrix}\;,
\end{equation}
we introduce variables:
\begin{eqnarray}
  z_k(s) &=& a_k(s) - a_{k+1}(s)\;, \quad \text{for}\quad k=1,\ldots,4, \\
  \beta_k(s) &=& b_k^2(s) \;,  \quad \text{for}\quad k=1,2,3, \\
  \delta_k(s) &=& d_k^2(s) \;,  \quad \text{for}\quad k=1,2, \\
  \gamma(s) &=& g^2(s) \;.
\end{eqnarray}
Then,  (\ref{general}) yields to
\begin{eqnarray}
  \phantom{w111}z_1' &=& 2\left(2\beta_1-\beta_2+\delta_1-\delta_2+\gamma\right)\;, \label{N4z1} \\
  \phantom{w111}z_2' &=& 2\left(-\beta_1+2\beta_2-\beta_3+\delta_1+\delta_2\right)\;, \label{N4z2} \\
  \phantom{w111}z_3' &=& 2\left(-\beta_2+2\beta_3-\delta_1+\delta_2+\gamma\right)\;, \label{N4z3} \\
  \left(\log \beta_1\right)' &=& -2z_1+4\sqrt{\frac{\beta_2\delta_1}{\beta_1}}+4\sqrt{\frac{\delta_2\gamma}{\beta_1}} \;, \label{N4b1} \\
  \left(\log \beta_2\right)' &=& -2z_2-4\sqrt{\frac{\beta_1\delta_1}{\beta_2}}+4\sqrt{\frac{\beta_3\delta_2}{\beta_2}} \;, \label{N4b2} \\
  \left(\log \beta_3\right)' &=& -2z_3-4\sqrt{\frac{\beta_2\delta_2}{\beta_3}}-4\sqrt{\frac{\delta_1\gamma}{\beta_3}} \;, \label{N4b3} \\
  \left(\log \delta_1\right)' &=& -2(z_1+z_2)+4\sqrt{\frac{\beta_3\gamma}{\delta_1}} \;, \label{N4d1} \\
  \left(\log \delta_2\right)' &=& -2(z_2+z_3)-4\sqrt{\frac{\beta_1\gamma}{\delta_2}} \;, \label{N4d2} \\
  \left(\log \gamma\right)' &=& -2(z_1+z_2+z_3)\;. \label{N4g} 
\end{eqnarray}
The advantage of the Mielke generator is that it allows to solve the Wegner equation step by step, where by steps we mean $k$-diagonals.
The tridiagonal case was solved in the previous section, then now we analyse the $5$-diagonal case, i.e. $g_0=0$ case.

\subsection{$g_0=0$ case}
Setting $g_0=0$ gives $\gamma(s)=0$ which leads to 5 equations (\ref{N4z1}), (\ref{N4z2}), (\ref{N4z3}), (\ref{N4d1}) and (\ref{N4d2}) with RHS being linear in functions $\beta_1$, $\beta_2$, $\beta_3$, $\delta_1$ and $\delta_2$. With the assumption that $a_i$, and then $z_i$, are in the form given by  (\ref{aN41}) -- (\ref{aN4N}), they can be solved yielding
\begin{eqnarray}
  \beta_1 &=& \frac{\widetilde{\eta}_1}{4\eta_1^2} - \frac{\eta_3}{\eta_1\eta_2} =  \frac{\widetilde{\rho_1}^2}{4\eta_1^2\eta_2} \;, \label{N4b1s} \\
  \beta_2 &=& \frac{\widetilde{\eta}_2}{4\eta_2^2} - \frac{\eta_3}{\eta_1\eta_2}
  - \frac{\eta_1}{\eta_2\eta_3} = \frac{\widetilde{\rho}_2^2}{4\eta_1\eta_2^2\eta_3} \;, \label{N4b2s} \\
  \beta_3 &=& \frac{\widetilde{\eta}_3}{4\eta_3^2} - \frac{\eta_1}{\eta_2\eta_3} = \frac{\widetilde{\rho}_3^2}{4\eta_2\eta_3^2} \;, \label{N4b3s} \\
  \delta_1 &=& \frac{\eta_3}{\eta_1\eta_2} \;, \label{N4d1s} \\
  \delta_2 &=& \frac{\eta_1}{\eta_2\eta_3} \;. \label{N4d2s}
\end{eqnarray}

As formerly, we look for functions $\eta_i$ in the form given by  (\ref{eta_k}), while  $\rho_1\in \mathcal{E}^4_1$, $\rho_2 \in \mathcal{E}^4_2$
and $\rho_3 \in \mathcal{E}^4_3$. The second equalities in  (\ref{N4b1s}) -- (\ref{N4b3s}) take into account that $b_1$, $b_2$ and $b_3$ may change a sign. Functions $\rho_i$ have to satisfy
\begin{eqnarray}
  \widetilde{\rho_1}^2 &=& \widetilde{\eta}_1\eta_2-4\eta_1\eta_3  \;, \label{N4r1} \\
  \widetilde{\rho_2}^2 &=& \eta_1\widetilde{\eta}_2\eta_3-4\eta_2(\eta_1^2 +\eta_3^2)  \;, \label{N4r2} \\
  \widetilde{\rho_3}^2 &=& \widetilde{\eta}_3\eta_2-4\eta_1\eta_3   \;. \label{N4r3}
\end{eqnarray}
In principle the above relations together with two of  (\ref{N4b1}) -- (\ref{N4b3}) allow to express e.g. all parameters of $\eta_1$, $\eta_3$, $\rho_1$, $\rho_2$ and $\rho_3$ by eigenvalues of $H_0$ and 5 of other $\eta_2$'s parameters. By comparing coefficients of exponential functions we get an overdetermined system of algebraic equations of the second and higher order for 19 unknowns. The eigenvalues of $H_0$ come from the forth degree equation, then we could expect that it should be possible to find all these parameters and even more, to find all parameters of $\eta_i$ in terms of $H_0$ entries, but it is left as an open question whether it is possible to manipulate all these algebraic equations without the necessity of solving equation of order higher than four.

\subsection{General case}
When $g_0\ne 0$ we still keep the assumption about $z_i$ given by (\ref{aN41}) -- (\ref{aN4N}),
but now, in the system of equations (\ref{N4z1}) -- (\ref{N4g}) there are only 4 equations with linear RHS for 6 functions $\beta_1$, $\beta_2$, $\beta_3$, $\delta_1$, $\delta_2$ and $\gamma$. From  (\ref{N4g}) we have
\begin{equation}
  \gamma =\frac{1}{\eta_1\eta_3} \label{gamma4}
\end{equation}
and from  (\ref{N4z1}) -- (\ref{N4z3}) we get relations between $\beta_1$, $\beta_2$, $\beta_3$, $\delta_1$ and $\delta_2$
\begin{eqnarray}
  \beta_1 &=& \frac{\widetilde{\eta}_1}{4\eta_1^2} - \frac{1}{\eta_1\eta_3} - \delta_1  \;, \label{N4bet1} \\
  \beta_2 &=& \frac{\widetilde{\eta}_2}{4\eta_2^2} - \frac{1}{\eta_1\eta_3} - \delta_1 - \delta_2  \;, \label{N4bet2} \\
  \beta_3 &=& \frac{\widetilde{\eta}_3}{4\eta_3^2} - \frac{1}{\eta_1\eta_3} - \delta_2  \;, \label{N4bet3}
\end{eqnarray}
but $\delta_1$ and $\delta_2$ are no longer given by  (\ref{N4d1s}) and  (\ref{N4d2s}).
The system of  (\ref{N4b1}) -- (\ref{N4d2}) can be considered
as the closed set of equations for
$\beta_1$, $\beta_2$, $\beta_3$, $\delta_1$ and $\delta_2$, and is the last obstacle on the way to solve the $4\times 4$ case completely
in addition to problems described in the former part. 
Numerical calculations exhibit an excellent compliance with formulas for $a_i(s)$ and $g(s)$ given accordingly by  (\ref{aN41}) -- (\ref{aN4N}) and (\ref{gamma4})
as well as with  (\ref{N4bet1}) -- (\ref{N4bet3}).
As previously, we expect that the solution to this system can be written
in terms of $\eta_1$, $\eta_2$ and $\eta_3$ and their derivatives.

\section{Connection to the matrix eigenvalue equation}
\label{eigen}
The Wegner equation describes a continuous unitary flow, then it conserves a number of quantities like eigenvalues or principal invariants $I_k$.
We have shown that the solution to the Wegner equation is built from functions $\eta_i$
being sums of exponential functions which exponents are bounded to matrix eigenvalues, but we can
expect further connections between $\eta_i$ and $H_0$. It turns out that for any $H\in \mathrm{Herm}(3, \mathbb{C})$ functions $\eta_1$ and $\eta_2$ may be expressed by eigenvectors of $H_0$ in the following way (we limit here to non-degenerate case):
\begin{eqnarray}
  \eta_1(s) &=& \frac{e^{u_1 s}}{|(v_1)_1|^2} +  \frac{e^{u_2 s}}{|(v_2)_1|^2} + \frac{e^{-(u_1+u_2) s}}{|(v_3)_1|^2} \;,  \\
\eta_2(s) &=& \frac{e^{-u_1 s}}{|(v_1)_3|^2} + \frac{e^{-u_2 s}}{|(v_2)_3|^2} + \frac{e^{(u_1+u_2) s}}{|(v_3)_3|^2} \;,
\end{eqnarray}
wherein $v_1$ is the eigenvector which corresponds to the largest eigenvalue and $v_3$ to the lowest, and where $(v_k)_l$ is the eigenvector $v_k$ with its $l$ component set to 1. If the first or third component is equal to 0 then a suitable change of basis can be done.
The proof is by a straightforward calculation.

The numerical observation reveals that the same holds in the case of $H\in \mathrm{Herm}(N, \mathbb{C})$ for functions $\eta_1$ and $\eta_{N-1}$, see (\ref{eta_k}).
Functions $\eta_1$ and $\eta_{N-1}$ allow to describe and confront with the numerical calculation the evolution of four matrix elements of $H(s)$, they are $H_{11}(s)$, $H_{NN}(s)$, $H_{1N}(s)$ and 
$H_{N1}(s)=H^*_{1N}(s)$. Proving this for $N=4$ should be in principle possible, while for $N\ge 5$, where roots cannot in general be expressed by radicals, can be more demanding. The open question is if similar relations exist for $\eta_k$, where $1<k< N-1$, for $N\ge 4$.

\section{\label{discuss}Summary}

Obtained results suggest that for any $H\in \mathrm{Herm}(N, \mathbb{C})$ the solution to the Wegner equation with the Mielke generator is composed of $N-1$ functions $\eta_i$ given by  (\ref{eta_k}). As there are no explicit formulas for roots of quintic equation and higher the parameters of $\eta_i$ also cannot in generally be found explicitly. The open question remains if these parameters can be found for every $H_0$ leading to a solvable eigenproblem.

We can also expect, based on the $2\times2$ results, that solutions in the case of other generators, such as Wegner or Brockett, will also be given in terms of functions being sums of exponential functions with exponents bounded to matrix eigenvalues, but the form of this dependence as well as the number of such functions and dependence of matrix elements on these functions will be different.

From the physical point of view we may look at the Wegner equation as a renormalization group driving the evolution of effective Hamiltonian with off-diagonal terms being coupling constants. Then,  properties of the effective Hamiltonian are not easily accessible. For $H\in \mathrm{Herm}(N, \mathbb{C})$ the evolution will be described by $2^N -2$ different exponential functions constituting $\eta_i$, even in the tridiagonal case. Thus, the exact evolution of 99 coupling constants may require computing more than $10^{30}$ exponential functions for only one value of~$s$. It means that 
the renormalization group, even if known exactly, can be practically applied only to not too large systems. Fortunately, in physically interesting cases there are at most a few coupling constants needed to be renormalized, but one can face such a problem when trying to apply the unitary flow to a discretized Hamiltonian. This means that a care must be preserved in obtaining approximated solutions, while the knowledge of exact solutions can provide invaluable help in their construction.

\appendix

\section{\label{c2x2}$2\times 2$ Hermitian matrix}

$2\times 2$ case can be presented in the same way as for $3\times 3$ case.
For $2\times 2$ Hermitian matrix
\begin{equation}
H(s) = \begin{bmatrix}
a(s) & g(s)\,e^{i\varphi(s)} \\
g(s)\,e^{-i\varphi(s)} & f(s) 
  \end{bmatrix}\;,
\end{equation}
the solution is given by
\begin{eqnarray}
  a(s)&=&\frac{Tr(H_0)}{2} + \frac{\eta'(s)}{2\eta(s)}\;,\\
  f(s)&=&\frac{Tr(H_0)}{2} - \frac{\eta'(s)}{2\eta(s)}\;,\\
  g(s)&=&\frac{\sgn(g_0)A}{\eta(s) }\;, \\
  \varphi(s)&=&\varphi_0 \;,
\end{eqnarray}
where
\begin{equation}
  \eta(s) = p \,e^{u s} + e^{-u s} 
\end{equation}
and 
\begin{eqnarray}
  p&=&\frac{2g_0^2+(a_0-f_0)^2+(a_0-f_0)\sqrt{(a_0-f_0)^2+4g_0^2}}{2g_0^2}\;,\\
  u&=&\sqrt{(a_0-f_0)^2+4g_0^2}\;,\\
  A&=&u\sqrt{p}\;.
\end{eqnarray}
We can also notice that  (\ref{eigen2exp}) is fulfilled.
%\begin{equation}
%  \{u, -u\} = \left\{2\left(r_1-\frac{Tr(H_0)}{2}\right), 2\left(r_2 -\frac{Tr(H_0)}{2}\right)\right\} \;,
%\end{equation}
%where $r_1$ and $r_2$
%\begin{eqnarray}
%  r_1&=&\frac{1}{2}\left[(a_0+f_0)+\sqrt{(a_0-f_0)^2+4g_0^2}\right]\;,\\
%  r_2&=&\frac{1}{2}\left[(a_0+f_0)-\sqrt{(a_0-f_0)^2+4g_0^2}\right]
%\end{eqnarray}
%are eigenvalues of $H_0$.

\section{\label{appR} Parameters of $\rho_1$ and $\rho_2$}

Assuming $\eta_1$ and $\eta_2$
in the form given by  (\ref{e1s}) and (\ref{e2s}), and $\rho_1$ and $\rho_2$ as
\begin{eqnarray}
\rho_1(s) &=& \sgn^{(1)}\left(q^{(1)}_1 \,e^{-u_1 s} + q^{(1)}_2 \,e^{-u_2 s} + q^{(1)}_3 \,e^{(u_1+u_2) s}\right) \;,  \label{rho1s}\\
\rho_2(s) &=& \sgn^{(2)}\left(q^{(2)}_1 \,e^{u_1 s} + q^{(2)}_2 \,e^{u_2 s} + q^{(2)}_3 \,e^{-(u_1+u_2) s}\right) \;.  \label{rho2s}
\end{eqnarray}
we can solve  (\ref{ro1-e}) and (\ref{ro2-e}) by comparing coefficients of exponential functions. This leads to
\begin{eqnarray}
  q^{(1)}_1 &=& \sgn^{(1)}_1 \left|u_1 + 2 u_2 \right| \sqrt{p^{(1)}_2 p^{(2)}_1} \;, \label{r11} \\
  q^{(1)}_2 &=& \sgn^{(1)}_2 \left|2u_1 + u_2 \right| \sqrt{p^{(1)}_1 p^{(2)}_2} \;,\\
  q^{(1)}_3 &=& \sgn^{(1)}_3 \left(u_1 - u_2 \right) \sqrt{p^{(1)}_1 p^{(1)}_2} \;,\\
  q^{(2)}_1 &=& \sgn^{(2)}_1 \left|u_1 + 2 u_2 \right| \sqrt{p^{(1)}_1 p^{(2)}_2} \;,\\
  q^{(2)}_2 &=& \sgn^{(2)}_2 \left|2u_1 + u_2 \right| \sqrt{p^{(1)}_2 p^{(2)}_1} \;,\\
  q^{(2)}_3 &=& \sgn^{(2)}_3 \left(u_1 - u_2 \right) \sqrt{p^{(2)}_1 p^{(2)}_2} \;. \label{r23}
\end{eqnarray}
A variety of possible combinations of signs in the initial conditions and solutions
for $q^{(i)}_j$ leads to seemingly complicated sign functions.
Functions $\sgn^{(i)}$ are given by
\begin{widetext}
\begin{eqnarray}
\sgn^{(1)} &=& \left\{
\begin{array}{ll}
  \sgn(b_0)\sgn\left(q^{(1)}_1 + q^{(1)}_2 + q^{(1)}_3 \right)  & \text{if } b_0 \ne 0 \;,\\
  1 & \text{if } b_0 = 0 \;,
\end{array} \right.\;,  \label{s1s}\\
\sgn^{(2)} &=& \left\{
\begin{array}{ll}
  1 & \text{if } b_0 \ne 0 \text{ and } g_0= 0 \;, \\
  \sgn(c_0)\sgn\left(q^{(2)}_1 + q^{(2)}_2 + q^{(2)}_3 \right) & \text{if } b_0 = 0 \;,
\end{array} \right.  \label{s2s}
\end{eqnarray}
and $\sgn^{(i)}_j$ by
\begin{eqnarray}
  \sgn^{(1)}_1 &=& \left\{
  \begin{array}{ll}
    \sgn\left((u_1 + 2 u_2)^2 + p^{(1)}_1 p^{(2)}_1 (u_1 - u_2)^2 - 4 A^2\right) & \text{if } b_0 \ne 0 \text{ and } g_0\ne 0\;,\\
    1 & \text{if } b_0 \ne 0 \text{ and } g_0=0 \;,\\
    \sgn(q^{(1)}_3)\sgn\left((u_1 + 2 u_2)^2 + p^{(1)}_1 p^{(2)}_1 (u_1 - u_2)^2 - 4 A^2\right) & \text{if } b_0 = 0 \;,
  \end{array} \right. \nonumber \\
  \label{sgn11} \\
  \sgn^{(1)}_2 &=& \left\{
  \begin{array}{ll}
    \sgn\left((2u_1 + u_2)^2 + p^{(1)}_2 p^{(2)}_2 (u_1 - u_2)^2 - 4 A^2\right) & \text{if } b_0 \ne 0  \text{ and } g_0\ne 0\;,\\
    1 & \text{if } b_0 \ne 0 \text{ and } g_0=0 \;,\\
    \sgn(q^{(1)}_3)\sgn\left((2u_1 + u_2)^2 + p^{(1)}_2 p^{(2)}_2 (u_1 - u_2)^2 - 4 A^2\right) & \text{if } b_0 = 0 \;,
    \end{array} \right.  \nonumber \\
  \label{sgn12} \\
  \sgn^{(1)}_3 &=& \left\{
  \begin{array}{ll}
    1 & \text{if } b_0 \ne 0 \;,\\
    \sgn(c_0)\sgn(g_0)\sgn\left(q^{(2)}_1 + q^{(2)}_2 + q^{(2)}_3 \right) \sgn\left(p^{(1)}_1 q^{(2)}_2 -p^{(1)}_2 q^{(2)}_1\right)& \text{if } b_0 = 0 \;,
    \end{array} \right.  \nonumber \\
  \label{sgn13} \\
  \sgn^{(2)}_1 &=& \left\{
  \begin{array}{ll}
    \sgn(q^{(2)}_3) \sgn\left((u_1 + 2 u_2)^2 + p^{(1)}_1 p^{(2)}_1 (u_1 - u_2)^2 - 4 A^2\right) & \text{if } b_0 \ne 0  \text{ and } g_0\ne 0\;,\\
    1 & \text{if } b_0 \ne 0 \text{ and } g_0=0 \;,\\
    \sgn\left((u_1 + 2 u_2)^2 + p^{(1)}_1 p^{(2)}_1 (u_1 - u_2)^2 - 4 A^2\right) & \text{if } b_0 = 0 \;,
    \end{array} \right.  \nonumber \\
  \label{sgn21} \\
  \sgn^{(2)}_2 &=& \left\{
  \begin{array}{ll}
    \sgn(q^{(2)}_3) \sgn\left((2u_1 + u_2)^2 + p^{(1)}_2 p^{(2)}_2 (u_1 - u_2)^2 - 4 A^2\right) & \text{if } b_0 \ne 0  \text{ and } g_0\ne 0\;,\\
    1 & \text{if } b_0 \ne 0 \text{ and } g_0=0 \;,\\
    \sgn\left((2u_1 + u_2)^2 + p^{(1)}_2 p^{(2)}_2 (u_1 - u_2)^2 - 4 A^2\right) & \text{if } b_0 = 0 \;,
    \end{array} \right.  \nonumber \\
  \label{sgn22} \\
  \sgn^{(2)}_3 &=& \left\{
  \begin{array}{ll}
    \sgn(b_0)\sgn(g_0)\sgn\left(q^{(1)}_1 + q^{(1)}_2 + q^{(1)}_3\right) \sgn\left(p^{(2)}_1 q^{(1)}_2-p^{(2)}_2q^{(1)}_1\right)& \text{if } b_0 \ne 0  \text{ and } g_0\ne 0\;, \\
    1 & \text{if } b_0 =0 \text{ or } g_0 = 0  \;.
    \end{array} \right. \nonumber \\
  \label{sgn23}
\end{eqnarray}
\end{widetext}

\section{\label{appA6}One degenerate eigenvalue}

If $u_1+2u_2=0$ or $2u_1+u_2=0$ then one eigenvalue of $H_0$ is degenerate. Formulas (\ref{aa}) -- (\ref{gg}) remain unchanged, only the 
functions $\eta_1$, $\eta_2$, $\rho_1$ and $\rho_2$ given by (\ref{e1tf}) -- (\ref{rho2t}) reduce to:
\begin{eqnarray}
  \eta_1(s) &=& p_{1} \,e^{u s} + e^{-2u s} \;,\label{eta1s} \\ 
  \eta_2(s) &=& p_{2} \,e^{-u s} + e^{2u s} \;,\label{eta2s} \\
  \rho_1(s) &=& q_{1} \,e^{-u s} \;,\label{rho1s2} \\
  \rho_2(s) &=& q_{2} \,e^{u s} \label{rho2s2}
\end{eqnarray}
where, see  (\ref{PP}) and (\ref{QQ}),
\begin{equation}
  u= -\frac{3P}{Q}  \label{u+}
\end{equation}
and
\begin{eqnarray}
  A &=&  \frac{3}{2}\,|u| \;, \label{As}\\
  p_{1} &=& -\frac{2 (3 a_0 - I_1 + 3 u)}{6 a_0 - 2 I_1 - 3 u} \;, \label{p1s}\\
  p_{2} &=& -\frac{2 (3 f_0 - I_1 + 3 u)}{6 f_0 - 2 I_1 - 3 u} \;,\\
  q_{1} &=& 2 b_0 (p_1 + 1) \sqrt{p_2 + 1} \;, \label{r1s} \\
  q_{2} &=& 2 c_0 (p_2 + 1) \sqrt{p_1 + 1} \;. \label{r2s}
\end{eqnarray}

\section{\label{appE1}Example of $3\times 3$ matrix}

We present here exact solutions to the Wegner equation with the initial condition being a symmetric
and a Hermitian matrix which differ only by phases.
In both cases we use the normalization of $\eta_1$, $\eta_2$ as $\eta_1(0)=\eta_2(0)=\sqrt{58/5}/3$, see the remark in the beginning of Sec. \ref{icon}. Coefficients of $\eta_i$ are given by (\ref{A}) -- (\ref{N22}) .

The symmetric case:\\
\begin{equation}
H_0^\mathrm{symm.}=\left[
\begin{array}{ccc}
 \frac{17}{6} & \sqrt{\frac{5}{87}} & 3 \sqrt{\frac{5}{58}} \\
 \sqrt{\frac{5}{87}} & \frac{547}{174} & \frac{26}{29} \sqrt{\frac{2}{3}} \\
 3 \sqrt{\frac{5}{58}} & \frac{26}{29} \sqrt{\frac{2}{3}} & \frac{350}{87} \\
\end{array}
\right]\;.
\end{equation}
The evolution of matrix elements are given by (\ref{aa}) -- (\ref{gg}), while
functions $\eta_1$, $\eta_2$, $\rho_1$ and $\rho_2$ are as follows (coefficients of $\rho_i$ come form (\ref{r11}) -- (\ref{r23}) and next are rescaled accordingly to the normalization of $\eta_i$):
\begin{eqnarray}
  \eta_1(s)&=&\frac{1}{9}\,\sqrt{\frac{29}{10}} \left(e^{3 s}+e^{-s}+4e^{-2 s} \right) \\
  &\approx& 0.1892\,e^{3 s} + 0.1892\,e^{-s} + 0.7569\, e^{-2 s} \;, \nonumber\\
\eta_2(s)&=&\frac{1}{9}\,\sqrt{\frac{10}{29}}\left(\frac{289}{25} \,e^{-3 s}+\,e^s+\frac{121}{25} \,e^{2 s}\right)  \\
  &\approx& 0.7543 \,e^{-3 s} + 0.06525\,e^s + 0.3158 \,e^{2 s}\;, \nonumber\\
\rho_1(s)&=& \frac{1}{135}\, \sqrt[4]{\frac{29\cdot2^3}{5}}\left(- 17 e^{-3 s} + 25 e^{s} + 22 e^{2 s} \right) \;,\\
\rho_2(s)&=& \frac{1}{27}\, \sqrt[4]{\frac{2}{29\cdot5^3}}\left( - 11 e^{3 s} + 187 e^{-s} + 136 e^{-2 s}\right) \;.
\end{eqnarray}

The Hermitian case:\\
\begin{equation}
H_0^\mathrm{Herm.}=\left[
\begin{array}{ccc}
 \frac{17}{6} & \sqrt{\frac{5}{87}}e^{\frac{i\pi}{3}} & 3 \sqrt{\frac{5}{58}}e^{-\frac{i\pi}{2}} \\
 \sqrt{\frac{5}{87}}e^{-\frac{i\pi}{3}} & \frac{547}{174} & \frac{26}{29} \sqrt{\frac{2}{3}}e^{-\frac{i\pi}{6}} \\
 3 \sqrt{\frac{5}{58}}e^{\frac{i\pi}{2}} & \frac{26}{29} \sqrt{\frac{2}{3}}e^{\frac{i\pi}{6}} & \frac{350}{87} \\
\end{array}
\right]\;,
\end{equation}
The evolution of matrix elements are given by (\ref{aa}) -- (\ref{ff}), (\ref{bherm}), (\ref{cherm}),  (\ref{hphib1i}) and (\ref{hphib2i}).
\begin{widetext}
%{\small
\begin{eqnarray}
\eta_1&=&\frac{2\sqrt{\frac{58}{15}}}{63 \left(1+2 \cos \frac{2 \alpha_1}{3}\right)}\left( 
  \left(-3 \sqrt{7} \cos \frac{\alpha_1}{3}+\sqrt{3} \left(3+7 \cos \frac{2 \alpha_1}{3}\right)\right) \, e^{u_1 s} \right. \\
&+&\sqrt{3}\csc \frac{\alpha_1}{3} \cos \frac{\pi -2 \alpha_1}{6} 
  \left(\sqrt{21} \sin \frac{\pi +2 \alpha_1}{6} +7 \sin \frac{\pi +4 \alpha_1}{6} -3-\sqrt{21} \cos \frac{\alpha_1}{3}
  \right)\,e^{ -(u_1+u_2)s} \nonumber\\
&+&\left. \csc \frac{\alpha_1}{3}\left(2 \cos \frac{\alpha_1}{3}-\sin \frac{\pi +2 \alpha_1}{6} \right)
  \left(3-7 \cos \frac{2 \alpha_1}{3}+\sqrt{21} \sin \frac{\pi +2 \alpha_1}{6} 
  +7 \sin \frac{\pi +4 \alpha _1}{6} \right)\, e^{u_2 s} \right) \nonumber \\
&\approx& 0.1625\, e^{2.796  s} + 0.3786\, e^{-0.3324  s} + 0.5942\, e^{-2.464 s}\;, \\
\nonumber\\
\eta_2&=&\frac{29}{45} \sqrt{\frac{203}{65267526}} \left( \left(240 \sqrt{7} \cos \frac{\alpha_1}{3}
+\sqrt{3} \left(419+406 \cos \frac{2 \alpha_1}{3}\right)\right) \sin \frac{\alpha_1}{3}\, e^{-u_1 s}  \right. \\
&+& \sqrt{3} \cos \frac{\pi -2 \alpha_1}{6}
\left(-419+80 \sqrt{21} \cos \frac{\alpha_1}{3}-80 \sqrt{21} \sin \frac{\pi +2 \alpha _1}{6}+406 \sin\frac{\pi +4 \alpha_1}{6} \right) e^{(u_1+u_2)s}
\nonumber \\
   &+& \left. \left(2 \cos \frac{\alpha_1}{3}-\sin \frac{\pi +2\alpha _1}{6}\right)
\left(419-80 \sqrt{21} \sin \frac{\pi +2 \alpha_1}{6}+406 \left(\sin \frac{\pi +4 \alpha_1}{6}-\cos \frac{2 \alpha_1}{3}\right)
\right) e^{-u_2 s} \right) \nonumber \\
&\approx& 0.8158\, e^{-2.796  s} + 0.03381\, e^{0.3324  s} + 0.2857\, e^{2.464 s} \;,
\end{eqnarray}
%}
\end{widetext}
where\begin{eqnarray}
\alpha_1&=&\arccos\left(\frac{2889}{5887} \sqrt{\frac{3}{7}} \right) \;,\\
u_1&=&2\sqrt{\frac{7}{3}} \cos \frac{\alpha _1}{3}  \;,\\
u_2&=&-2\sqrt{\frac{7}{3}} \sin \left(\frac{\pi}{6} + \frac{\alpha _1}{3} \right) \;.
\end{eqnarray}
\begin{widetext}
\begin{figure}
\includegraphics[width=0.99\columnwidth]{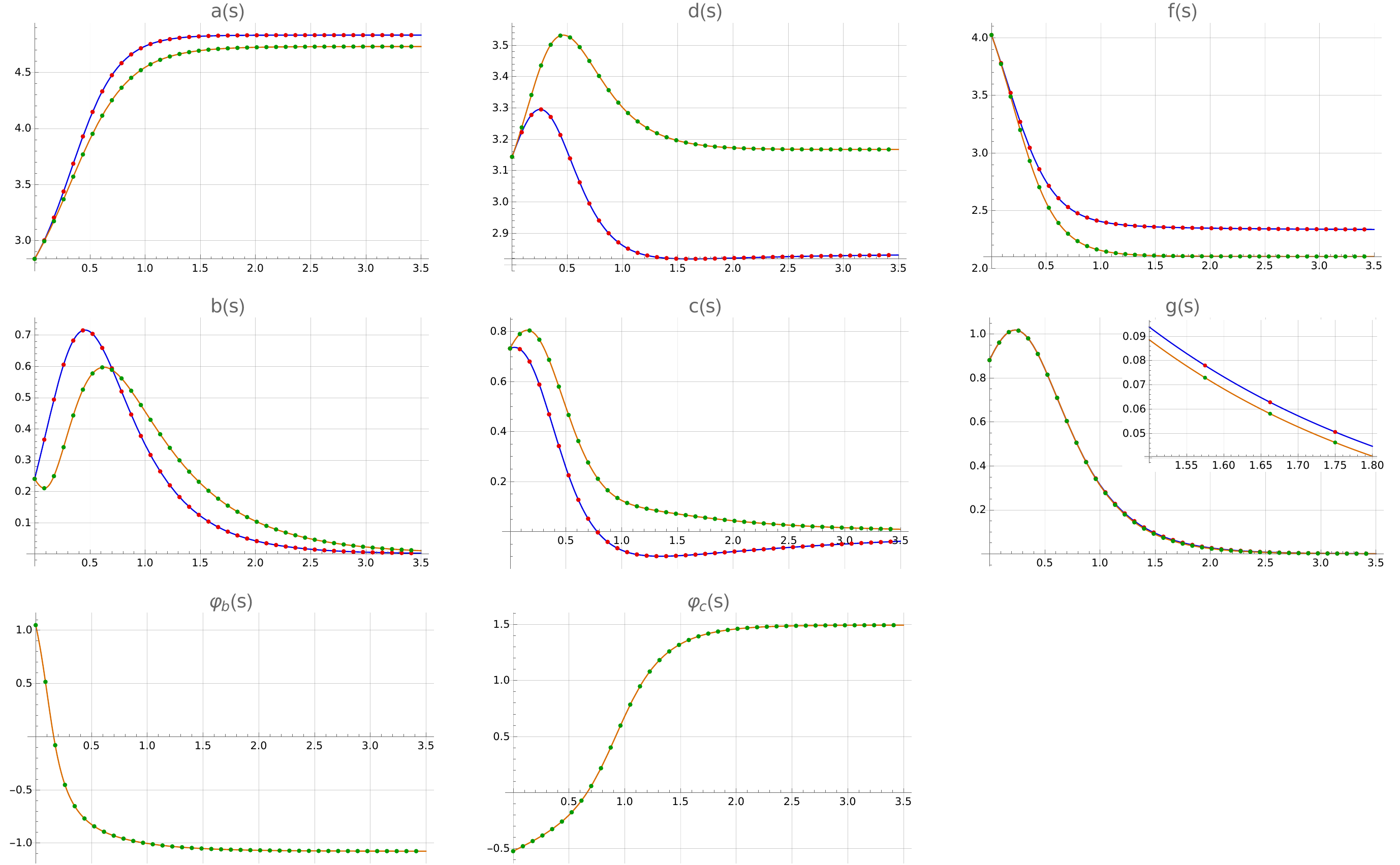}
\caption{The exact solution (solid lines) plotted together with the 4th order Runge-Kutta numerical calculations (dots).
Colours blue and red represent results for the symmetric matrix, orange and green for Hermitian. The inset in the $g(s)$ plot shows that results for the symmetric and Hermitian matrix are different. In the case of symmetric matrix plots of off-diagonal elements represent matrix elements evolution, while for the Hermitian matrix the evolution of absolute values of matrix elements (off-diagonal). The last two plots show functions $\varphi_b(s)$ and $\varphi_c(s)$.}
\label{figure}
\end{figure}
\end{widetext}

\section{\label{appE2}Example of the $5\times 5$ tridiagonal matrix}

As there are no general expressions for roots of quintic equation in terms of radicals
the example is obtained by choosing the parameters in $\eta_1$ to fulfil the condition (\ref{FF}). This fixes all other parameters in $\eta_2$, $\eta_3$ and $\eta_4$.
We choose $\Tr(H_0)=5$ and exponents $u_i$ in  (\ref{eta_k}) to be equal to $-4,\; -1/2,\; 1,\; 3/2,\; 2$ what fixes the eigenvalues of $H_0$ to $-1,\; 3/4,\; 3/2,\; 7/4,\; 2$, see  (\ref{eigen2exp}).
Next, with the help of  (\ref{p2sol}) -- (\ref{pNsol}) we get four $\eta$ functions and then  (\ref{aN41}) - (\ref{betatriN-1}) give $a_i(s)$ and $b_i(s)$.
Finally, by taking $s=0$ we get $H_0$.
\begin{widetext}
\begin{equation}
H_0=\left[
\begin{array}{ccccc}
 \frac{1740683}{3678812} & \frac{385 \sqrt{\frac{76252037}{2}}}{1839406} & 0 & 0 & 0 \\
 \frac{385 \sqrt{\frac{76252037}{2}}}{1839406} & \frac{22231005067381}{70129227185011} & \frac{6 \sqrt{57988393428551}}{76252037} & 0 & 0 \\
 0 & \frac{6 \sqrt{57988393428551}}{76252037} & \frac{39263366363260747}{38462269852632232} & \frac{3 \sqrt{4323173670686265}}{504409736} & 0 \\
 0 & 0 & \frac{3 \sqrt{4323173670686265}}{504409736} & \frac{8664841726526959}{5719587241749384} & \frac{154 \sqrt{126102434}}{11339169} \\
 0 & 0 & 0 & \frac{154 \sqrt{126102434}}{11339169} & \frac{18982507}{11339169} 
\end{array}
\right]\;.
\end{equation}
Functions $\eta_1$, $\eta_2$, $\eta_3$, $\eta_4$ are:
%{\small
\begin{eqnarray}
  \eta_1 &=& \frac{512 e^{-4 s}}{225}+\frac{8 e^{-s/2}}{9}+\frac{16 e^s}{25}+\frac{32 e^{3 s/2}}{49}+\frac{128 e^{2 s}}{121} \;, \\
  \eta_2 &=& \frac{12544 e^{-9 s/2}}{2025}+\frac{2048 e^{-3 s}}{225}+\frac{123904 e^{-5 s/2}}{11025}+\frac{65536 e^{-2 s}}{3025}+\frac{8 e^{s/2}}{25}+\frac{256 e^s}{441}+\frac{1600 e^{3s/2}}{1089} \nonumber \\
  && +\frac{32 e^{5 s/2}}{1225}+\frac{512 e^{3 s}}{3025}+\frac{256 e^{7 s/2}}{5929} \;, \\
  \eta_3 &=& \frac{3136 e^{-7 s/2}}{225}+\frac{61952 e^{-3 s}}{2025}+\frac{100352 e^{-5 s/2}}{1089}+\frac{30976 e^{-3 s/2}}{11025}+\frac{65536 e^{-s}}{3025}+\frac{8192 e^{-s/2}}{1225} \nonumber \\
  && +\frac{16 e^{2 s}}{1225}+\frac{16 e^{5 s/2}}{121}+\frac{3200 e^{3 s}}{53361}+\frac{64 e^{9 s/2}}{148225} \;, \\
  \eta_4 &=& \frac{968 e^{-2 s}}{225}+\frac{6272 e^{-3 s/2}}{121}+\frac{256 e^{-s}}{9}+\frac{512 e^{s/2}}{1225}+\frac{2 e^{4 s}}{5929} \;.
\end{eqnarray}
%}
\begin{figure}
\includegraphics[width=0.99\columnwidth]{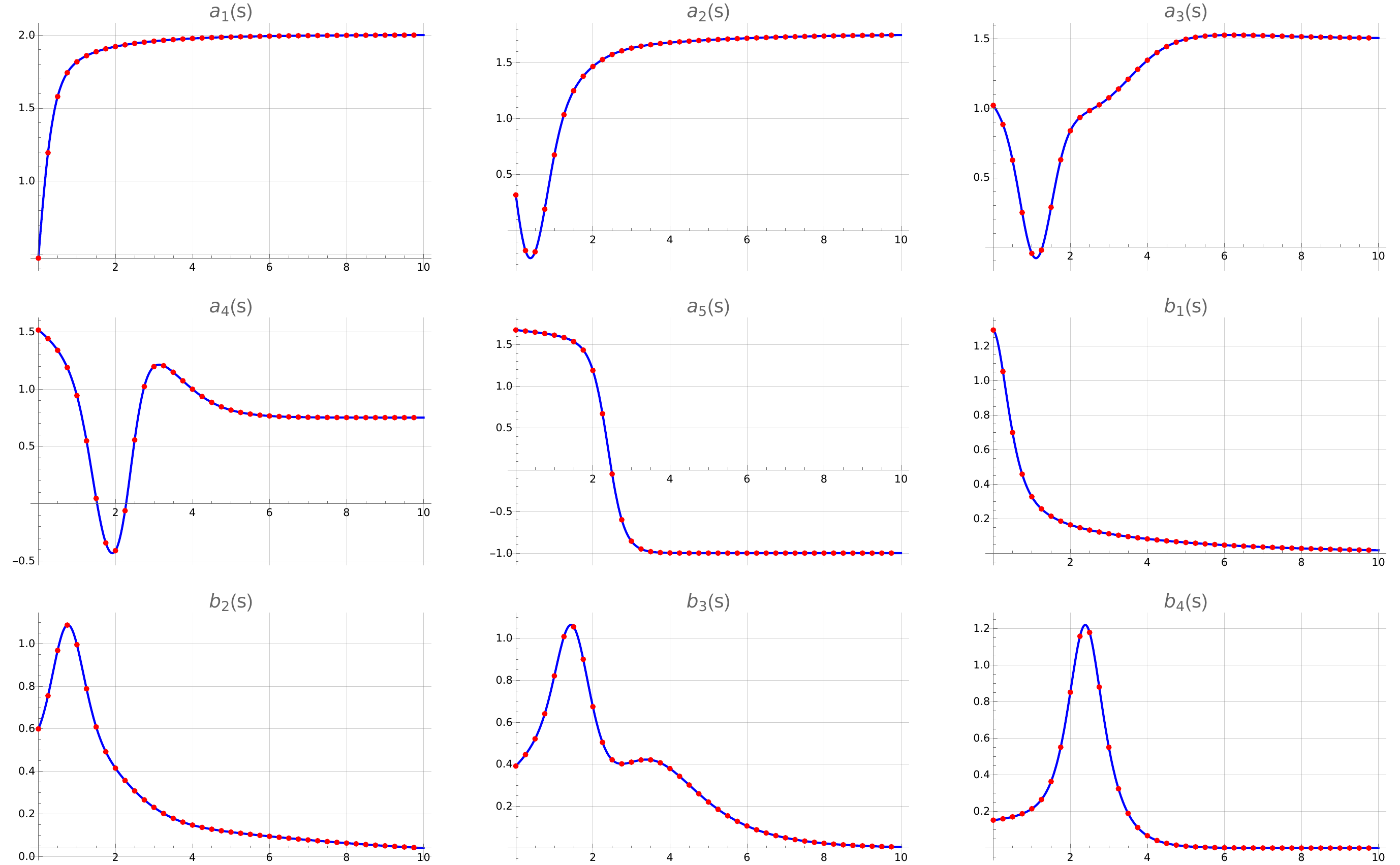}
\caption{The exact solution (solid blue lines) plotted together with the 4th order Runge-Kutta numerical calculations (red dots).}
\label{figure2}
\end{figure}
\end{widetext}

\nocite{*}
\bibliography{library}

\end{document}